\documentclass[
prd,
twocolumn,showpacs,superscriptaddress,groupedaddress]{revtex4}  
\usepackage[dvipdfmx]{graphicx}
\usepackage{amsmath,amssymb,amsxtra}
\usepackage{graphicx}  
\usepackage{dcolumn}   
\usepackage{bm}        
\usepackage{amssymb}   
\usepackage{float,color} 

\newcommand{\alf}{Alfv\'en\ }
\newcommand{\alfv}{Alfv\'enic\ }
\newcommand{\BZ}{BZ\ }
\newcommand{\pa}{\partial }

\newcommand{\al}{\alpha }
\makeatletter
\def\vmargin@#1#2#3{
\setbox0=\hbox{#3}%
\rule[#1]{0pt}{\ht0}%
\lower\dp0\hbox{\rule[-#2]{0pt}{\dp0}}%
\box0%
}
\def\vmargin#1#2#3{
\mathchoice
{\vmargin@{#1}{#2}{$\displaystyle #3$}}
{\vmargin@{#1}{#2}{$\textstyle #3$}}
{\vmargin@{#1}{#2}{$\scriptstyle #3$}}
{\vmargin@{#1}{#2}{$\scriptscriptstyle #3$}}
}
\makeatother
\begin{document}

\widetext


\title{Blandford-Znajek process as Alfv\'enic superradiance}
%
\author{Sousuke Noda}
\email{sousuke.noda@yukawa.kyoto-u.ac.jp}
\affiliation{%
  Center for Gravitation and Cosmology, College of Physical Science and Technology, Yangzhou University, Yangzhou 225009, China\\}%
\affiliation{%
Yukawa Institute for Theoretical Physics, Kyoto University,
Kyoto 606-8502, Japan
}%

\author{Yasusada Nambu}%
\email{nambu@gravity.phys.nagoya-u.ac.jp}
 \affiliation{%
  Department of Physics, Graduate School of Science, Nagoya University,
  Chikusa, Nagoya 464-8602, Japan}
\author{Takuma Tsukamoto}
\email{tsukamoto.takuma@h.mbox.nagoya-u.ac.jp}
 \affiliation{%
  Department of Physics, Graduate School of Science, Nagoya University,
  Chikusa, Nagoya 464-8602, Japan}
\author{Masaaki Takahashi}
\email{mtakahas@auecc.aichi-edu.ac.jp}
\affiliation{Department of Physics and Astronomy, Aichi University of
  Education, Kariya, Aichi 448-8542, Japan}


\begin{abstract} 
 The superradiant scattering of \alf waves (\alfv superradiance) in a forcefree magnetosphere is 
 discussed to reveal the relationship between the Blandford-Znajek (BZ) process and superradiance.
For simplicity, we consider a four-dimensional rotating black string spacetime 
of which each $z=\text{const}$ slice is a Ba\~nados-Teitelboim-Zanelli solution 
 as an analogy of the equatorial plane of the Kerr spacetime.
 Then, it is confirmed that the condition for \alfv superradiance coincides 
  with that for the BZ process, and the wave amplification can be very large due to a 
  resonant scattering for some parameter sets of the wave frequency and the angular velocity 
  of the magnetic field line.
Moreover, by analysis of the Poynting flux, 
we first show that the BZ process can be interpreted as the long wavelength limit of \alfv superradiance. 
\end{abstract}


\maketitle

\section{Introduction}
  As rotational energy extraction processes from black holes, the Penrose process, 
  superradiance, and  the Blandford-Znajek (BZ)  process are widely discussed.  
  The Penrose process is energy exchange between particles by splitting or collisions 
  inside the ergoregion \cite{Penrose, Wagh1985}. 
  By transitioning one particle to a negative energy orbit, the other particle can acquire energy 
  larger than that of the initial incident particle. Superradiance is a similar mechanism for waves  
  \cite{Zeldovich1971,Zeldovich1972,Starobinsky1973,Starobinsky1974,Lasota2014,Brito2015}. 
  The waves incident toward the black hole are scattered, and they can be propagated to a distant 
  region with amplification if the following condition is satisfied: 
   $0<\omega/m < \Omega_\text{H}$, where $\Omega_\text{H}$ is the angular velocity 
   of the black hole, $m$ is the azimuthal quantum number for a wave 
   mode, and $\omega$ is the frequency of the incident wave. 

   The BZ process \cite{Blandford1977} is an energy extraction mechanism via 
   electromagnetic fields from a rotating black hole. It is thought that the electromagnetic fields 
   in the vicinity of black holes are so strong that they are dominant and 
   the inertia of plasma can be ignored (forcefree approximation). 
   Therefore, the BZ process is often discussed for the forcefree magnetosphere. 
   The mechanism works 
   as follows. 
   The magnetic torque acts on magnetic field lines due to the spacetime dragging effect, 
   and the rotational energy of spacetime is transported outward 
   in the form of the Poynting flux. 
   This energy extraction is possible under the condition 
   $0 < \Omega_F <  \Omega_\text{H}$, where $\Omega_F$ is the angular velocity of 
   the magnetic field lines. 
  The BZ process has been studied for several situations in analytical way 
  \cite{Toma2014,Toma2016,Jacobson2019,Kinoshita2018} and by numerical calculations 
\cite{Komissarov2004,Komissarov2005,Koide2006,McKinney2006,Ruiz2012,Koide2014,Koide2018} 
for black hole magnetospheres.
 Toma and Takahara \cite{Toma2014} revealed that the ergoregion is crucial for generating the 
 outward Poynting flux, and Kinoshita and Igata \cite{Kinoshita2018} discussed that the light 
 surface of the background magnetic field has to be inside the ergoregion for the BZ process. 
 Moreover, there are several works regarding the relationship between 
 the Penrose process or superradiance and the BZ process \cite{Komissarov2009,Lasota2014}.
 However, the relationship has not been clarified so far.

 The \BZ process is driven by background electromagnetic fields, but in a forcefree 
 magnetosphere, propagation of fast magnetosonic waves 
 and \alf waves also occur. 
 Thus, these waves can contribute to the energy extraction process, for example, 
 via superradiance. 
 Indeed, superradiance for fast magnetosonic waves which is longitudinal mode has been discussed 
 in papers by Uchida \cite{Uchida1997b,Uchida1997c} and van Putten \cite{Putten1999}. 
 The condition for it is the same as the ordinary superradiance for scalar, vector, and tensor waves.
 Furthermore, it was argued that superradiance for \alf waves (Alfv\'enic superradiance) 
 does not occur through the discussion based on the eikonal approximation. 
 However, it is still possible to amplify \alf waves in the treatment without eikonal 
 approximation. Indeed, in the numerical calculations \cite{Koide2006, Komissarov2004}, 
 the outward propagation of \alf waves generated in the ergoregion is 
 important for energy extraction. Since an \alf wave is a transverse wave mode propagating 
 along magnetic field lines due to the magnetic tension, we can discuss energy extraction 
 along magnetic field lines if \alfv superradiance is possible. 
 To see this, we analyze the wave equation for \alf waves. 
 Moreover, by decomposition of the Poynting flux into the contribution of the background 
 electromagnetic field and the perturbation, it will be shown that the BZ process is 
 explained as the long wavelength limit of \alfv superradiance.
 
In order to obtain a magnetosphere solution around a black hole, it is necessary to solve 
the general relativistic Grad-Shafranov equation \cite{Blandford1977}. 
For the Kerr spacetime, this equation cannot be solved globally in an analytical way.
Therefore, in this paper, we consider a simpler geometry with cylindrical symmetry which can be a good model to discuss the essence of phenomena in the Kerr spacetime.

This paper is organized as follows. In Sec.~II, we derive a stationary and axisymmetric 
magnetosphere solution in the cylindrical spacetime, and the BZ process in this spacetime is discussed. Then, we give a perturbation to the magnetosphere to obtain the wave equations in section III. Sec.~IV is devoted to the derivation of the condition for \alfv superradiance and 
the evaluation of how much the \alf waves can be amplified. Sec~V discusses the relationship between the BZ 
process and \alfv superradiance before concluding the paper in Sec~VI.

\section{Background magnetosphere solution}
\subsection{Black cylinder spacetime}
We consider the forcefree electromagnetic fields in a four-dimensional black 
string spacetime (black cylinder) \cite{Jacobson2019} with a scale factor $f(z)$ 
as a benchmark to discuss the BZ process.
The metric $g_{\lambda \nu}$ is given as
\begin{equation}
ds^2=-\alpha^2dt^2+\alpha^{-2}dr^2 +r^2\left( d\varphi-\Omega dt \right)^2  + f(z)^2dz^2,
\label{eq:cylinder}
\end{equation}
where $\alpha$ and $\Omega$ are functions of the radial coordinate given as 
$\alpha^2:={(r^2-r_+^2)(r^2-r_-^2)}/{(r^2 \ell^2)}$, 
$\Omega:={r_+ r_-}/{(r^2 \ell)}$, 
and $\ell$ denotes the AdS curvature scale related to the negative cosmological 
constant as $\Lambda_3=-\ell^{-2}$. This spacetime has two horizons as the Kerr spacetime and their radii $r_\pm$ are given by $\alpha(r_\pm)=0$. 
Each constant-$z$ slice of the spacetime is a Ba\~nados-Teitelboim-Zanelli 
black hole \cite{BTZ1992}, and hence, the horizon geometry is cylindrical. 
The mass and angular momentum of the black cylinder can be written with $r_\pm$ 
as $M={(r_+^2+r_-^2)}/{\ell^2}, J={2r_+ r_-}/{\ell}$. These parameters satisfy $J\leq M\ell$,
and hence the spin parameter defined as $a:=J/(\ell M)$ should be less than unity for 
the spacetimes to have horizons.
Using the parameter $a$, the angular velocity at the horizon $\Omega_\text{H}:=\Omega(r_+)$ is 
\begin{equation}
\Omega_\text{H}=\dfrac{1}{\ell}\left(\dfrac{a}{1+\sqrt{1-a^2}}\right).
\label{eq:OmH}
\end{equation}
The reason why we added the ``extra'' dimension to the three-dimensional black hole 
solution is that we need to consider a four-dimensional spacetime to discuss the ordinary electromagnetic fields for astrophysics. Moreover, the Grad-Shafranov equation can be solved
by choosing the functional form of the scale factor $f(z)$ properly.
Since, in this model, $f(z)$ is an arbitrary function of $z$, we choose it as $f(z)=\cos{(\mu z)}$ 
for $\mu^2>0$ and $f(z)=\cosh{(|\mu|z)}$ for $\mu^2<0$ with a constant $\mu$.
\subsection{Forcefree magnetosphere in the black cylinder spacetime}
We consider a stationary and axisymmetric forcefree magnetosphere in this spacetime. 
Within the forcefree approximation mentioned in Sec. I, the Maxwell 
equation yields the following set of equations: $F_{\lambda \nu}\nabla_{\beta} F^{\nu \beta}=0,\ \nabla_{[\lambda} F_{\nu \beta]}=0$.
The field strength $F_{\lambda \nu}$ satisfying these equations can be represented by two 
scalars, called Euler potentials \cite{Uchida1997,Gralla2014}, as 
\begin{equation}
F_{\mu\nu}=\partial_{\mu}\phi_1\partial_{\nu}\phi_2-\partial_{\nu}\phi_1\partial_{\mu}\phi_2,
\label{eq:F}
\end{equation}
and the Maxwell equation reduces to the equations for $\phi_1$ and $\phi_2$:
\begin{equation}
\partial_{\lambda} \phi_i \partial_{\nu} \left[\sqrt{-g}\left(g^{\lambda \alpha} g^{\nu \beta}-g^{\nu \alpha}g^{\lambda \beta}\right)\partial_{\alpha}\phi_1 \partial_{\beta} \phi_2\right]=0,\ \ i=1,2,
\label{eq:basic_eq}
\end{equation}
where $\lambda, \nu, \alpha, \beta=t, r, \varphi, z$.
For the stationary and axisymmetric solution, we can consider the following ansatz for 
Euler potentials \cite{Uchida1997a}: $\phi_1=\Psi(z),\ \ \phi_2=h(r)+ \varphi-\Omega_{F} t$, 
where the angular velocity of the magnetic field lines $\Omega_F$ is a constant.
 From Eq. \eqref{eq:basic_eq}, we obtain 
\begin{equation}
\phi_1=-\psi_z \int dz f(z) ,\ \ \phi_2=\dfrac{I}{2\pi \psi_z }\int  \dfrac{dr}{r \alpha^2}+ \varphi-\Omega_{F} t,
\label{eq:Euler}
\end{equation}
where constants $\psi_z$ and $I$ are the magnetic monopole line density 
located on the $z$ axis and the electric current, respectively. 
The function $\phi_1$ corresponds to the stream function of the magnetosphere and 
$\phi_1=\text{const}$ gives the so-called magnetic surface. 
For the present model, a magnetic surface is a constant-$z$ plane, whereas 
$\phi_2=\text{const}$ defines the configuration of the magnetic field lines on 
each magnetic surface \cite{Uchida1997,Uchida1997a,Uchida1997b,Uchida1997c,Gralla2014,Kinoshita2018}. 

\begin{figure}[H] 
  \centering
  \includegraphics[width=6.cm]{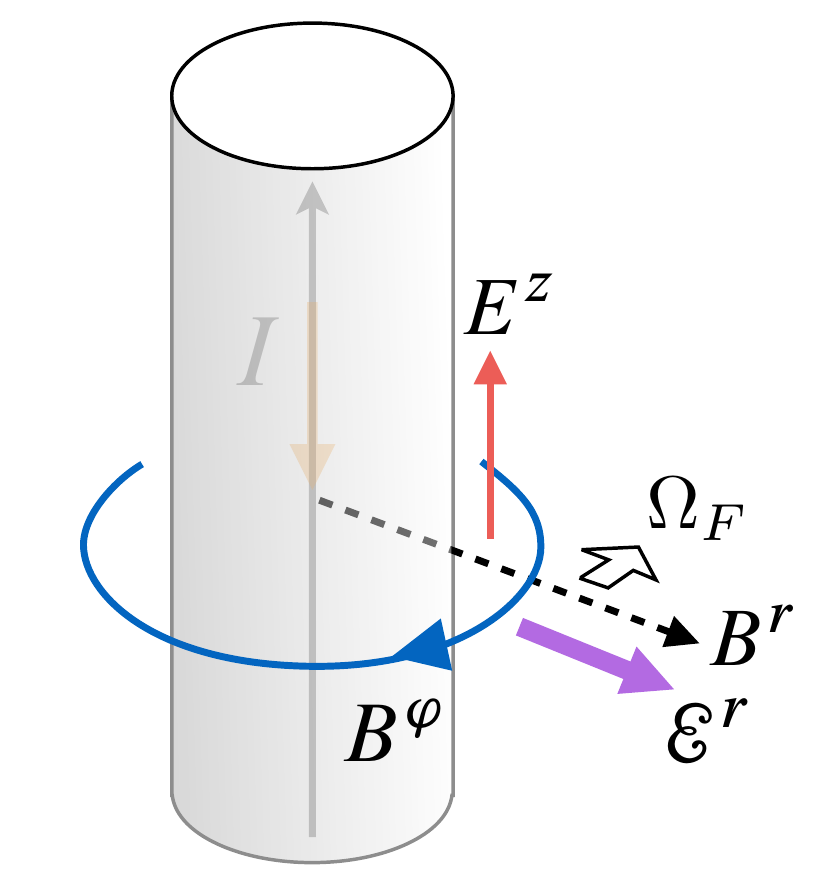}
  \caption{\footnotesize{The configuration of the electromagnetic fields in the black cylinder 
  spacetime. The grey cylinder represents the horizon. Note that for $I>0$ case, the current flows in the $-z$ direction}
    }
    \label{fig:configuration}
\end{figure}

\begin{figure}[H] 
  \centering
  \includegraphics[width=6.cm]{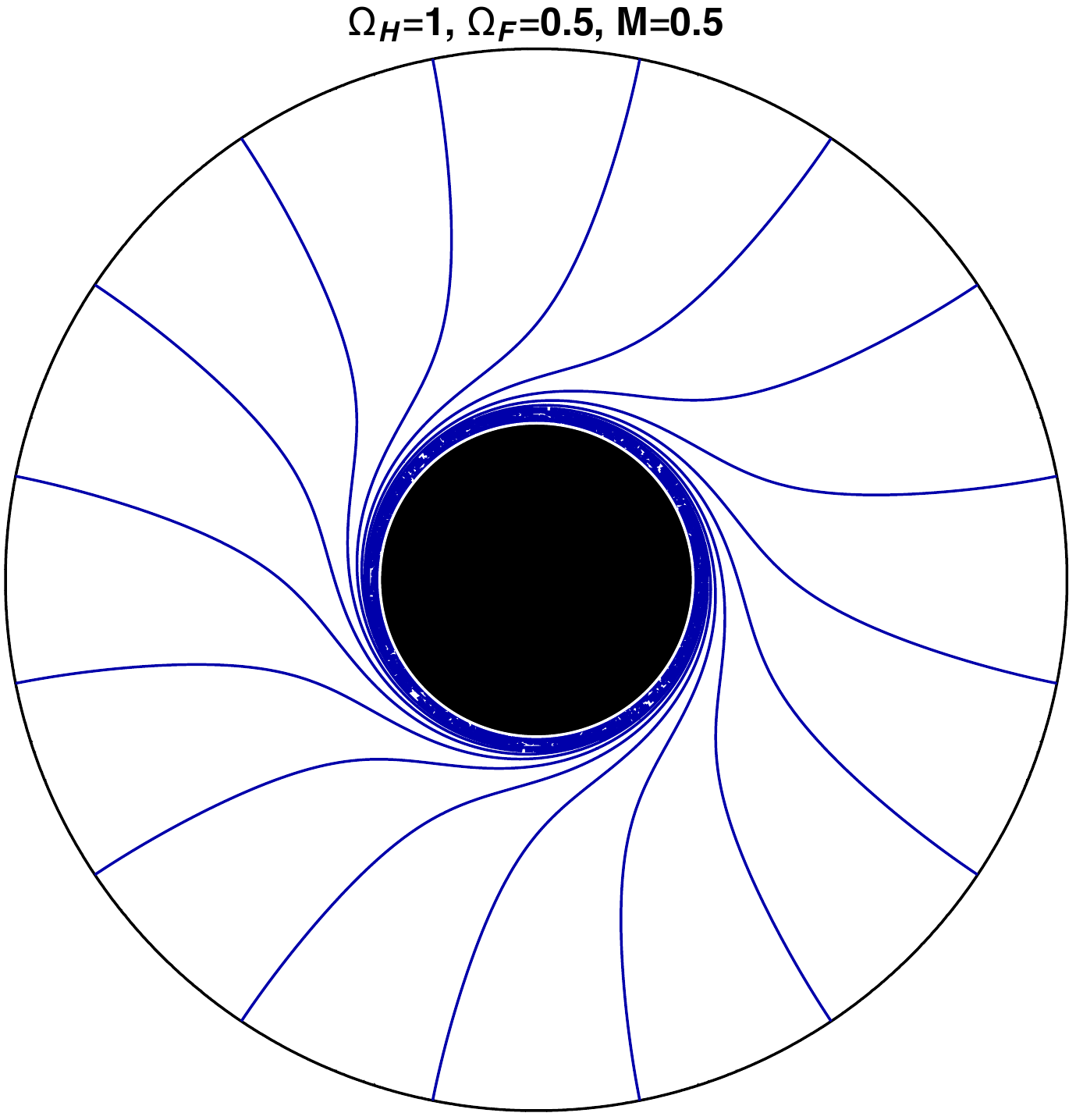}
  \caption{\footnotesize{The snapshot ($t=\text{const}$) of the magnetic field 
  lines ($\phi_2=\text{const}$) on a magnetic surface (z=\text{const}).  
  The white circle is the black hole horizon 
  and the outer circle represents the AdS boundary. For the present parameters, 
  the radius of the light surface is $r_\text{LS}\simeq 1.1 r_+$.
  For the radial coordinate, we mapped the range $r_+<r<\infty$ to the finite one 
  $\arctan{(r_+/\ell)}<\tilde{r}<\pi/2$ through the transformation $\tilde{r}=\arctan{(r/\ell)}$.
  }}
\label{fig:fieldline}
\end{figure}

To clarify the situation we are considering, we compute the components of the 
electromagnetic fields measured by a fiducial observer of which four-velocity is given as 
$u_{\nu}=(-\alpha,0,0,0)$. The electric and magnetic fields are defined as 
$E^{\nu}=F^{\nu\beta}u_{\beta}$ and $B^{\nu}=-{}^*F^{\nu \beta}u_{\beta}$, 
respectively. The dual tensor is defined as ${}^*F^{\nu \beta}=-\epsilon^{\nu \beta \lambda \rho }/(2\sqrt{-g})F_{\lambda \rho}$ with the completely antisymmetric tensor.
Substituting the solution \eqref{eq:Euler} into these definitions, we get the following nonzero components of the electromagnetic fields:
\begin{equation}
  E^z=\frac{\psi_zf(z)}{\al}(\Omega_{F}-\Omega),\ 
  B^r=\frac{\al}{r}\psi_zf(z),\ 
  B^\varphi=-\frac{f(z)I}{2\pi r^2\al}.
  \label{eq:components}
\end{equation}
The axial current $I$ generates the toroidal magnetic field
  $B^\varphi$ (Amp\`{e}re's law), and the rotating (moving) radial magnetic field 
  $B^r$ (sourced by a magnetic monopole density distributed on the $z$-axis) generates the electric field 
  $E^z$ (Faraday's law).
The configuration and the magnetic field lines for the present system are shown in 
Fig. \ref{fig:configuration} and Fig. \ref{fig:fieldline}, respectively.
\subsection{The BZ process for the present model}
The BZ process works for this model as discussed by Jacobson 
and Rodriguez \cite{Jacobson2019} who considered $f(z)=1$ case. 
  In the present model, $E^z$ and $B^\varphi$ generate the radial Poynting flux
  $\mathcal{E}^r$.
Although the detailed computation of the Poynting flux including the wave effect 
(perturbation) will be discussed in Sec. V, let us now show only 
the flux by the background magnetosphere here:
\begin{equation}
 {\cal{E}}^r= I \Omega_F \psi_z,
\label{eq:BZ_flux}
\end{equation}
where we evaluated the flux flowing through a short section of the cylinder with radius $r$ and 
the unit $z$-length in the vicinity of the magnetic surface at $z=0$. 
The sign of the current $I$ determines that of $\mathcal{E}^r$.
Since the regularity of the electromagnetic field at the horizon
requires the following relation called the Znajek condition:  
\begin{equation}
I=2\pi r_+ \psi_z (\Omega_\text{H}- \Omega_{F}),
\label{eq:Znajek}
\end{equation}
the Poynting flux becomes outward when the inequality 
\begin{equation}
0<\Omega_F<\Omega_\text{H},
\end{equation}
is satisfied. 
Namely, the rotational energy of the black hole is extracted if the black hole horizon rotates 
faster than the magnetic field line.
\section{Wave propagation}
\subsection{Wave equations and wave modes}
Let us discuss the propagation of waves in the background magnetosphere. 
First of all, we define the perturbation to the Euler potential 
$\phi_i\rightarrow \phi_i+\delta \phi_i(t, r, \varphi, z)$ as 
$\delta \phi_i := \zeta_{i}^{\lambda} \partial_{\lambda} \phi_i$.
The displacement vectors $\zeta_i^{\lambda}$ are assumed to be functions of $t, r, \varphi$.
Hereafter, we focus on the wave propagations on the magnetic 
surface given by $z=0$, where the scale factor $f(z)$ is unity,  
its first derivative becomes zero, and the second derivative is $-\mu^2$. 
Taking the first-order terms of Eq. \eqref{eq:basic_eq}, 
we obtain the following equations for $\delta \phi_1$ and $\delta \phi_2$:
\begin{align}
&\partial_\nu \phi_2 \partial_{\lambda}\left(\sqrt{-g}\partial^{[ \lambda}\delta\phi_1 \partial^{\nu]}\phi_2\right)=0,\label{eq:alf_perturbation}\\
&\partial_j\left(\sqrt{-g}\partial^j \delta \phi_2 \right)=0,
\label{eq:fast}
\end{align}
where $j=t, r, \varphi$ and the square bracket represents the anticommutator. 
The perturbation $\delta \phi_2$ obeys the Klein-Gordon equation, and 
the dispersion relation is the same as that of a massless particle. 
This is one of the features of the fast magnetosonic wave \cite{Uchida1997b,Uchida1997c}.
Although the fast magnetosonic wave propagates on a magnetic surface due 
to the assumption of the perturbation, in general, its propagation is not restricted on a magnetic surface
\footnote{For the Kerr case, we can explicitly show that the fast magnetosonic waves can propagate 
in the off-magnetic surface direction under the symmetric assumption on the perturbation.},
whereas the perturbation $\delta \phi_1$ corresponds to the \alf wave, which always propagates along a magnetic field line on a magnetic surface, as we will see later. 
  It can be shown that the Poynting flux of the BZ process flows on the 
magnetic surface \cite{Kinoshita2018}, and our aim is to investigate the relationship between 
the BZ process and the propagation of the \alf waves. 
Therefore, we focus only on the \alf wave mode.

\subsection{Propagation of \alf waves}
 Considering the similarity between the propagation of \alf waves and the 
 Poynting flux via the BZ process, we discuss the propagation of 
 \alf waves on the magnetic surface $z=0$. We first rewrite Eq. \eqref{eq:alf_perturbation} 
 in terms of a parameter along a magnetic field line $\sigma$ and the time coordinate for a 
 corotating observer of the magnetic field line $\tau$. 
 The coordinates $(\tau, \sigma, \rho)$ are introduced through the following transformation: 
\begin{equation}
 t=\tau,\ r=\sigma,\ \varphi=\rho -\dfrac{I}{2\pi \psi_z}\int \dfrac{d\sigma }{\sigma \alpha^2}+\Omega_{F} \tau,
\end{equation}
where $\rho$ is $\phi_2$ itself, and each $\rho=\text{const}$ gives a 
magnetic field line. Therefore, $\rho$ is a coordinate perpendicular to the 
magnetic field lines. The differential operators with respect to the new coordinates are 
$\partial_\tau=\partial_t+\Omega_F \partial_{\varphi}$ and 
$\partial_\sigma  =\partial_r-I/(2\pi \psi_z  r \alpha^2)\partial_\varphi$. 
 In these coordinates, the second equation of \eqref{eq:alf_perturbation} yields 
 \begin{align}
  &-C_1(\delta\phi_1)_{\tau\tau}-\al^2 \sigma \left[\dfrac{\Gamma}{\sigma}
  \left(\pa_\sigma-\dfrac{I \sigma (\Omega-\Omega_F)}{2\pi \psi_z \Gamma \al^2}\pa_\tau\right)\delta \phi_1 \right]_{\sigma}
    \notag \\
  & +\dfrac{I \sigma(\Omega-\Omega_F)}{2\pi \psi_z}(\delta \phi_1)_{\tau \sigma}
  +\sigma^2 \al^2 C_2 (\delta \phi_1)_{zz}
=0,
    \label{eq:del1}
\end{align}
where $(\delta \phi_1)_{zz}=-\mu^2 \delta \phi_1$ due to the definition of the 
perturbation and the background field configuration. 
The functions $C_1$ and $C_2$ are defined as $C_1:=1+{I^2}/{(4\pi^2\al^2\psi_z ^2)}$ and 
 $C_2:={I^2}/{(4\pi^2\sigma^2 \al^2 \psi_z ^2)}-{(\Omega-\Omega_{F})^2/\al^2}+1/\sigma^2$, 
 respectively. 
 The function $\Gamma$ is the norm of the corotating vector of the field line
$\chi_F^\nu=\left(\pa_t\right)^\nu+\Omega_F \left(\pa_\varphi \right)^\nu$: 
 \begin{equation}
\Gamma=g_{\lambda \nu }\chi_F^{\lambda} \chi_F^{\nu}=-\al^2+r^2(\Omega-\Omega_{F})^2.
 \end{equation}
The zero point of $\Gamma$ gives the location of the light surface, which is 
the causal boundary for \alf waves \cite{Gralla2014}, and we denote its location by 
$r=r_\text{LS}$. 
For black hole magnetospheres, in general, there exist inner and outer 
light surfaces. The inner one is caused by the gravitational redshift, whereas the 
outer one stems from the fact that the velocity of rigidly rotating magnetic field lines 
exceed the speed of light.
For the present model, there is only one light surface in the vicinity of 
the black cylinder's horizon due to the asymptotic feature of the spacetime, and 
the norm is negative everywhere outside the light surface 
\footnote{This is one of the different points from the magnetosphere in the Kerr spacetime, for which 
there exists the outer light surface as well.}.
Note that Eq. \eqref{eq:del1} does not have a derivative term with respect to $\rho$. 
This means the perturbation $\delta \phi_1$ propagates only on a two-dimensional 
sheet spanned by $\tau$ and $\sigma$, called a field sheet \cite{Uchida1997,Uchida1997a,Uchida1997b,Uchida1997c,Gralla2014,Kinoshita2018}, 
which represents the time evolution of a magnetic field line. 
Therefore, we can identify $\delta \phi_1$ as an \alf wave.
 Of course, $\delta \phi_1$ has $\rho$ dependence through the function $A(\rho)$ as 
 $\delta \phi_1 \propto A(\rho)$. However, this factor is a constant for wave propagation 
 along a magnetic field line. 

\begin{figure}[H] 
  \centering
  \includegraphics[width=8.6cm]{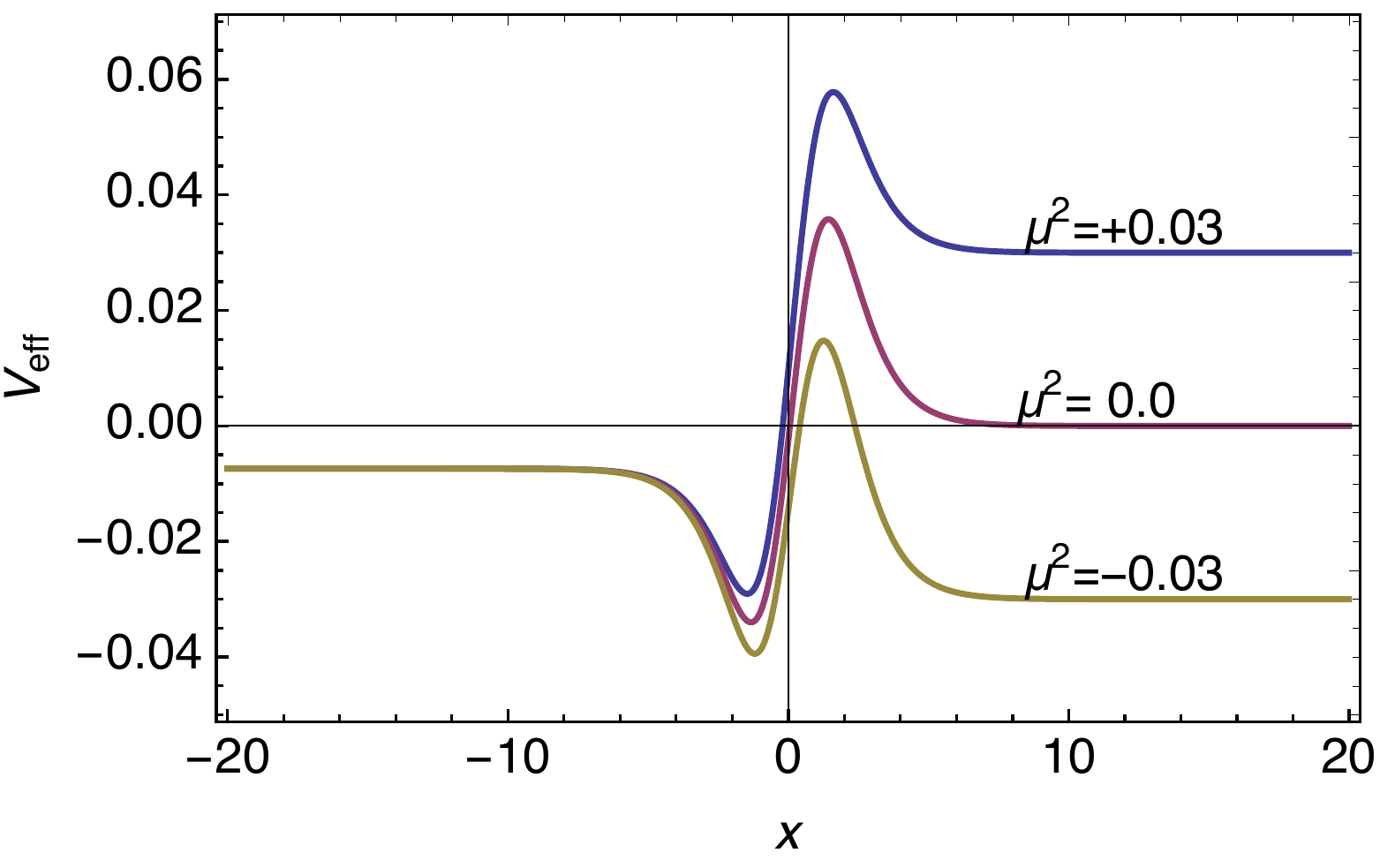}
  \caption{\footnotesize{$V_\text{eff}$ with $a=0.9,\ \Omega_F\ell=0.5,\ \omega \ell=0.1$ for 
  $\mu^2 \ell^2= 0.03$, 
  $\mu^2 \ell^2= 0$, and $\mu^2 \ell^2= -0.03$. 
  The light surface is located at $x=-\infty$ in this coordinate.
   \alf waves can propagate to the distant region only for $\mu^2<0$.}
    }
    \label{fig:potential}
\end{figure}

 To eliminate the cross term of $\tau$ and $\sigma$, we choose another set of 
 coordinates $(T, X)$ on the field sheet, defined as 
\begin{equation}
 \tau=-\dfrac{I}{2 \pi \psi_z}\int dX X \dfrac{\Omega- \Omega_F}{\al^2 \Gamma}+T,\ \sigma=X.
\end{equation}
 $\pa_T=\pa_\tau$ and 
 $\pa_X=\pa_\sigma-{I \sigma (\Omega-\Omega_{F})}/{(2\pi\al^2\Gamma\psi_z) }\pa_\tau$. 
We can separate the variables as 
$\delta \phi_1 = R(X)A(\rho)e^{-i\omega T} \partial_z \phi_1$ on $z=0$ plane, 
then Eq. \eqref{eq:del1} yields
\begin{equation}
 -\Gamma X \pa_X\left(\frac{\Gamma}{X}\pa_XR\right)
+VR=0,
\label{eq:wave_Alfven_tausigma}
\end{equation}
where
\begin{equation} 
V:=\dfrac{\omega^2}{\al^2}
    \left[ C_1\Gamma-
    \dfrac{I^2X^2(\Omega-\Omega_{F})^2}{(4\pi^2\al^2\psi_z ^2)}\right]
-\mu^2\Gamma C_2 X^2.
\end{equation} 
In the present treatment, we assume $0<\Omega_{F} \ell <1$ for which
the region $X<X_\text{LS}$ becomes a super-\alf region as in the case 
of the ordinary context of a black hole magnetosphere \cite{Beskin}.
Since $\Gamma$ can be factorized as $\Gamma=-(\gamma/\ell^2)(X^2-X_\text{LS}^2)$ with 
$\gamma:=(1-\ell^2\Omega_{F}^2)$, we introduce the dimensionless ``tortoise'' coordinate
 $x$ as 
 \begin{equation}
 \dfrac{d}{dx}:=(X-X_\text{LS})\dfrac{d}{dX}, \quad (-\infty <x<+\infty).
 \end{equation}
 In this coordinate, the position of the light surface is $x=-\infty$.
Then, introducing a new wave function defined by the relation $R=K^{-1/2}\tilde
R$, $K:=1+{X_\text{LS}}/{(X_\text{LS}+\ell\, e^x)}$, Eq. \eqref{eq:wave_Alfven_tausigma} 
can be written in the form of the Schr\" odinger equation:
\begin{equation}
 -\tilde R_{xx}+V_\text{eff} \tilde R=0,\ \  V_\text{eff}:=\frac{K_{xx}}{2K}-\frac{K_x^2}{4K^2}+\frac{V\ell^4}{\gamma^2
        X^2K^2},
       \label{eq:wave_Alfven_TX}
\end{equation}
where $X=X_\text{LS}+\ell\, e^x$.
The asymptotic form of the effective potential is 
\begin{equation}
  V_\text{eff}\sim 
  \begin{cases}
 \quad  \quad  \quad  \quad  \quad \mu^2 \ell^2
   &\ \text{for}\ x\rightarrow +\infty \\
    -\dfrac{\omega^2\ell^4r_{+}^2}{4\gamma^2\al^4}(\Omega_\text{H}
    -\Omega_{F})^2(\Omega-\Omega_{F})^2&\ \text{for}\ x\rightarrow-\infty.
  \end{cases}
  \label{eq:as_veff}
\end{equation}
We show the behavior of the effective potential for several values of $\mu^2$ in Fig. \ref{fig:potential}. 
For $\mu^2<0$, in the short wavelength limit 
($\omega^2 \ell^2 \gg 1$, $|\mu^2| \ell^2 \gg 1$), 
there is no reflection of waves  because the top of the potential 
barrier goes below zero, whereas for $\mu^2 \geq 0$, the waves are confined in a finite region $x<0$ due to the 
potential barrier in the $x>0$ region.
 \noindent
 We focus only on the $\mu^2<0$ case to discuss \alfv superradiance because in the case of 
 $\mu^2\geq 0$, there is no outward propagation to a distant region from the black cylinder. 

\section{\alfv superradiance}
\begin{figure}[H] 
  \centering
  \includegraphics[width=8.6cm]{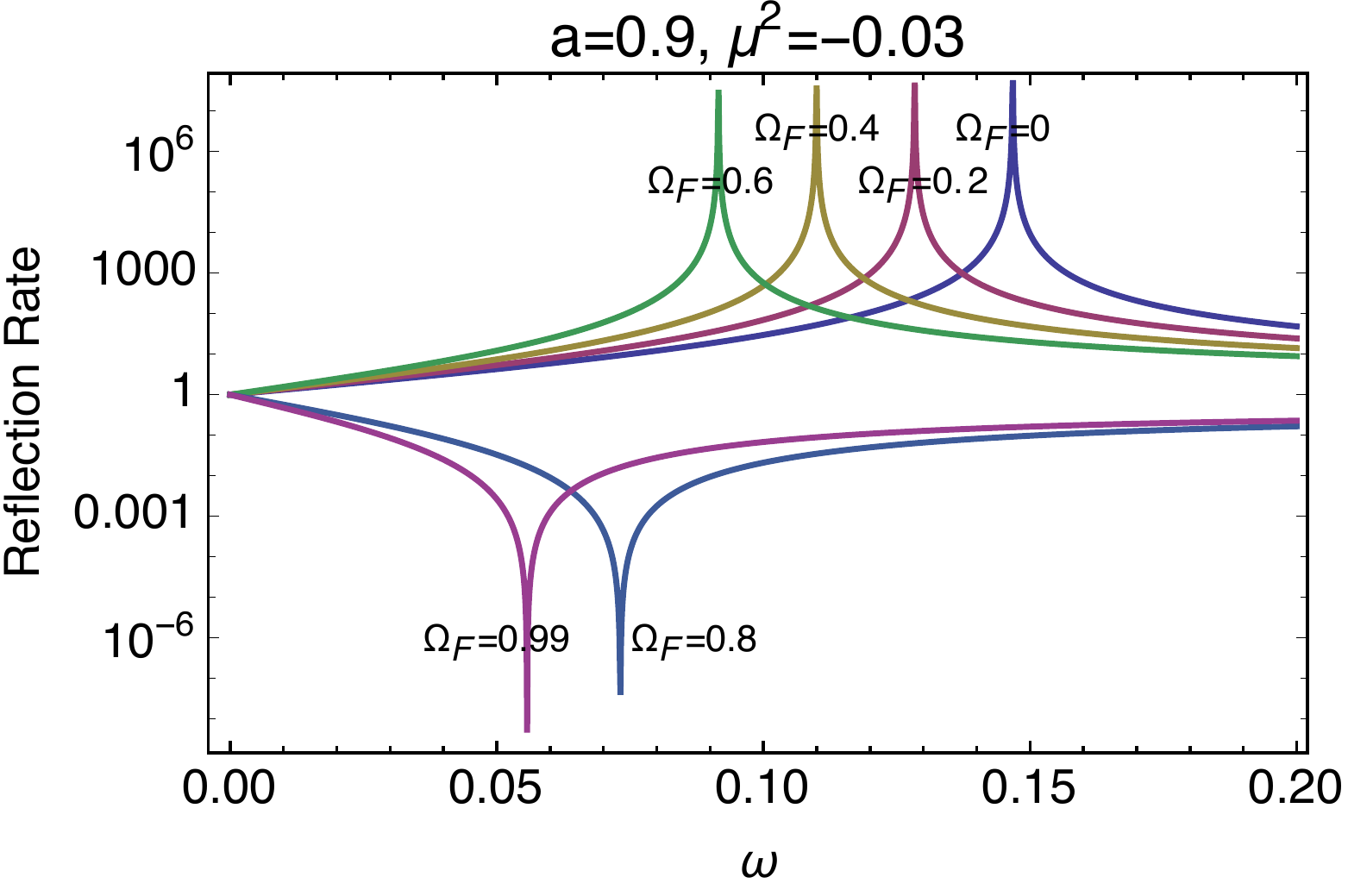}
  \caption{\footnotesize{The reflection rate of the \alf waves for several $\Omega_F$.}
    }
    \label{fig:refrate1}
\end{figure}

\begin{figure}[H] 
  \centering
  \includegraphics[width=8.6cm]{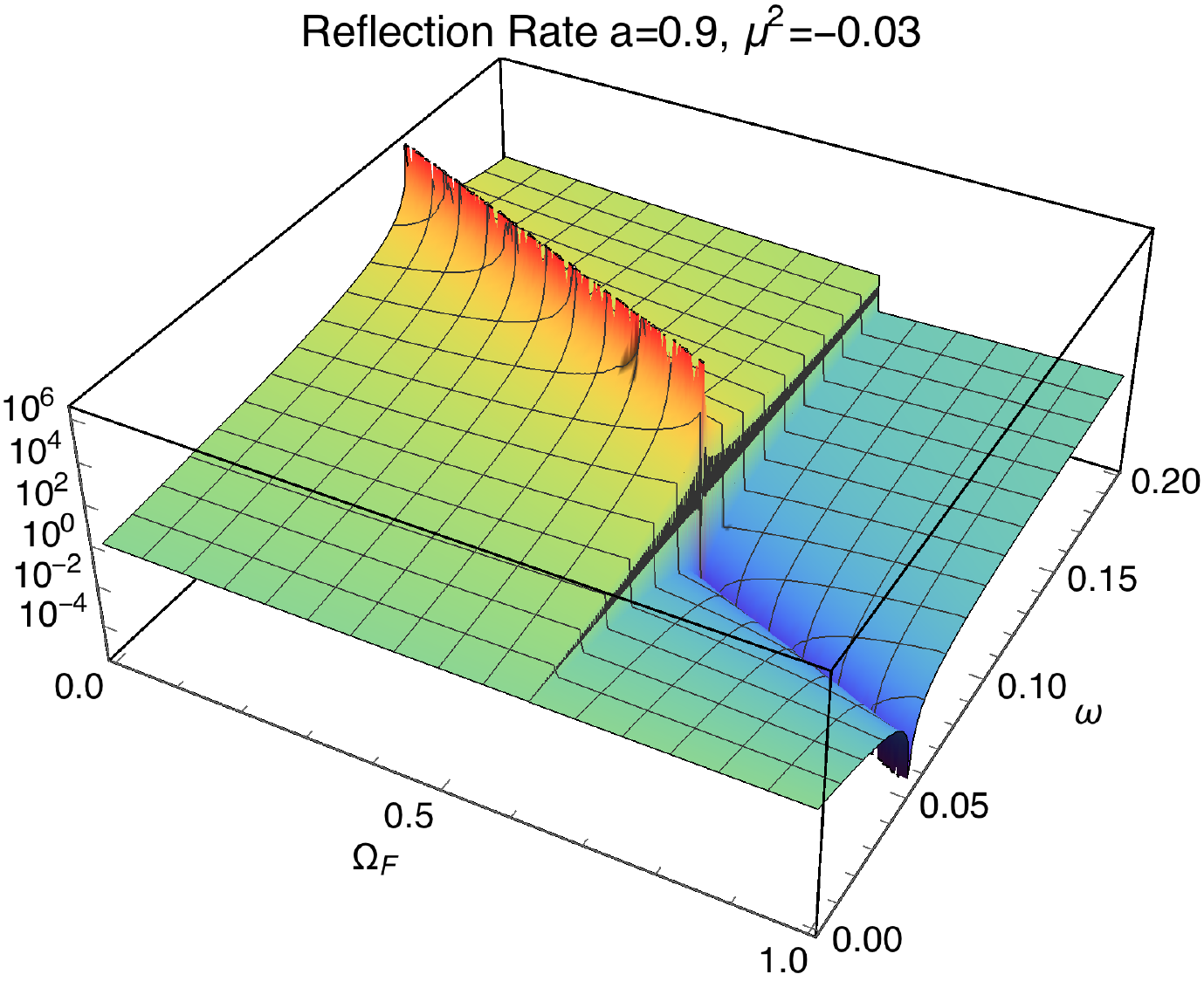}
  \caption{\footnotesize{3D Plot of the reflection rate on the $\omega$-$\Omega_F$ plane.}}
  \label{fig:refrate2}
\end{figure}

Since the light surface is the causal boundary for \alf waves, we can write the asymptotic solutions with the proper definition of ingoing mode in the vicinity of $X=X_\text{LS}$ as follows:
\begin{equation}
\tilde R\sim 
 \begin{cases}
 A_\text{in}e^{- i\sqrt{-\mu^2}\ell x}
 +A_\text{out}e^{ i\sqrt{-\mu^2}\ell x}
  &{\makebox[0.5em][r]{\text{for}}\, x\rightarrow +\infty} \\
\exp\left[- i\dfrac{\omega \ell^2 r_{+}}{2\gamma }|\Omega_\text{H}-\Omega_{F}|{\displaystyle \int
   dx\frac{\Omega_\text{F}-\Omega}{\al^2}}  \right] 
   &\makebox[0.5em][r]{\text{for}} \, x\rightarrow-\infty,
 \end{cases}
 \label{eq:asympt_func}
 \end{equation} 
where $A_\text{in}$ and $A_\text{out}$ are the coefficients of the ingoing mode and the 
outgoing mode, respectively \footnote{Strictly speaking, the wave does not propagate at $x\rightarrow \infty$ 
because the exponents do not include the frequency $\omega$. 
At a distant point, $\omega$-dependence of the effective potential  is 
$V_\text{eff}=\left[ \mu^2-\omega^2 /(\gamma e^{2x})\right]\ell^2.$ 
Therefore, we define the ingoing/outgoing modes by considering the sign of 
the second term including $\omega$. However, it is very small at a distant point, and hence we omit this term in the asymptotic form of the wave function \eqref{eq:asympt_func}. 
}.
Note that the absolute value symbol and positivity of $\Omega_\text{H}$ and $\Omega_F$ 
are necessary to define the ingoing mode properly for 
both the $0<\Omega_F< \Omega_\text{H}$ and $0< \Omega_\text{H} <\Omega_F$ cases.
The conservation of the Wronskian at the light surface and the infinity gives the following reflection rate:
\begin{equation}
\left|\dfrac{A_\text{out}}{A_\text{in}}\right|^2=1- \dfrac{\omega \ell r_{+}|\Omega_\text{H}-\Omega_{F}|}{2\gamma \alpha_{\text{LS}}^2 \sqrt{-\mu^2}}\dfrac{\Omega_{F}-\Omega_\text{LS}}{|A_\text{in}|^2},
\label{eq:refrate}
 \end{equation} 
where $\alpha_\text{LS}:=\alpha(r_\text{LS})$ and $\Omega_\text{LS}:=\Omega(r_\text{LS})$.
If the following inequality is satisfied,
\begin{equation}
 0< \Omega_{F}<\Omega_\text{LS},
 \label{eq:alfcond}
\end{equation}
then the reflection rate $|A_\text{out}/A_\text{in}|^2$ exceeds unity, namely, the \alf wave is 
 amplified when scattered by the potential (\alfv superradiance). 
  Note that condition \eqref{eq:alfcond} is different from the superradiant condition for 
  ordinary waves (e.g. scalar waves) $0 <\omega/m < \Omega_\text{H}$. 
  In the case of \alf waves, the condition \eqref{eq:alfcond} depends on the angular velocity of the magnetic field lines $\Omega_F$ instead of on $\omega/m$. This reflects the fact that an 
  \alf wave propagates along a magnetic field line and the separation of variable $\varphi$ 
  is not necessary. Furthermore, the upper boundary of the condition 
  \eqref{eq:alfcond} is the angular velocity of the spacetime at the light surface 
  instead of that of the horizon because the light surface is a one-way 
boundary for \alf waves.
Although condition \eqref{eq:alfcond} does not have the wave frequency, 
the reflection rate $|A_\text{out}/A_\text{in}|^2$ itself depends on 
$\omega$, as we show in Fig. \ref{fig:refrate1} and Fig. \ref{fig:refrate2}.
%
%
\noindent 
As shown in Fig. \ref{fig:refrate1} and Fig. \ref{fig:refrate2}, indeed, the reflection rates exceed unity if the \alfv superradiant condition is satisfied. 
The value of the upper bound of the condition \eqref{eq:alfcond} is 
$\Omega_\text{LS}\simeq 0.63$ for $a=0.9$.
Moreover, we observed that the reflection rate becomes very large or very small for 
 some parameter sets $(\omega, \Omega_F)$. 
 These features correspond to resonant scattering and perfect absorption 
 of \alf waves. They occur when the values of the effective 
potential at the light surface and far region coincide with each other.
From the asymptotic values of the effective potential \eqref{eq:as_veff}, the resonant 
frequency $\omega_\text{res}$ is obtained as 
$\omega_\text{res}=(r_+/\ell)\sqrt{-\mu^2}\left(1-\ell^2 \Omega_\text{H} \Omega_F\right)$.

\section{\alfv superradiance and the BZ process}
How does \alfv superradiance relate to the BZ process? 
Interestingly, it turns out that condition \eqref{eq:alfcond} is exactly the 
same as the condition for the BZ process $0<\Omega_F < \Omega_\text{H}$ as follows: 
Considering the fact that $\Omega_\text{LS}$ is a function of $\Omega_F$:
\begin{equation}
\Omega_\text{LS}(\Omega_F)=\dfrac{1-\ell^2 \Omega_F^2}{2\ell(1-\ell \Omega_F a)}a,
\label{eq:Om_LS}
\end{equation}
we solve the inequality \eqref{eq:alfcond} for $\Omega_F$.
Then, we obtain 
\begin{equation}
0< \Omega_F <  \ell^{-1} \left({a}/({1+\sqrt{1-a^2}})\right)=\Omega_\text{H},
\label{eq:Alf_BZ}
\end{equation}
where the equality comes from Eq. \eqref{eq:OmH}.

We investigate the Poynting flux, including the effect of \alf waves. 
To do that, we introduce the conserved energy flux vector with the timelike Killing vector 
$({\partial_{t}})^{\nu}$ as ${P}^{\lambda}=-T^{\lambda}_{\ \nu}\left(\partial_{t}\right)^{\nu}$,
where the energy momentum tensor is 
$T_{\lambda\nu}=F_{\lambda \alpha}F_{\nu}^{\ \alpha}-({1}/{4})F_{\alpha \beta}F^{\alpha \beta}g_{\lambda\nu}$.
Integrate $P^{\lambda}$ for the azimuthal angle $\varphi$ and define ${\cal{E}}^{\lambda}:=2\pi r P^{\lambda}$, then the energy flux per unit time over a section of a cylinder with a unit $z$-length and a constant radius $r \simeq r_\text{LS}$ is 
\begin{equation}
 {\cal{E}}^r=
I \Omega_F \psi_z\Bigl[\underbrace{\vmargin{0ex}{1.6ex}{1}}_{\text{BZ}} 
\underbrace{\overbrace{-\dfrac{\mu^2 }{2}|A|^2|R|^2}^{\text{zero mode}}+O(\omega^2)}_{\text{perturbation}}\ \Bigr].
\label{eq:Pr}
\end{equation}
Note that all the terms have the common factor 
$I\Omega_F \propto \Omega_F (\Omega_\text{H}-\Omega_F)$. 
This factor for the BZ term comes from the Znajek condition, whereas the Poynting flux of the perturbation is proportional to $(\Omega_\text{LS}- \Omega_F)$, which comes 
 from the condition for \alfv superradiance.
However, it can be shown that 
$(\Omega_\text{LS}-\Omega_F) \propto (\Omega_\text{H}-\Omega_F)$, 
therefore we can factorize Eq. \eqref{eq:Pr} with $\Omega_F (\Omega_\text{H}-\Omega_F)$.
The perturbation term depending on $\omega^2$ enhances the flux of the BZ process 
when \alfv superradiance occurs. 
Furthermore, the zero mode of the perturbation enhances the flux for the $\mu^2<0$ case 
in which \alf waves can propagate to a distant region.
Actually, the contribution of the zero mode term can be incorporated 
into the BZ term as a small deformation of the background field: 
$\psi_z^2 \rightarrow \psi_z^2 \left(1-({\mu^2}/{2})|A|^2 |R|^2\right)$. 
If we redefine the modified one as a new background field \footnote{Note that redefining the background field by shifting $\psi_z^2$ is consistent with 
the components of the electromagnetic fields \eqref{eq:components} and the Znajek condition \eqref{eq:Znajek}.}, Eq. \eqref{eq:Pr} with the 
limit $\omega \rightarrow 0$ is nothing but the energy flux of the BZ process for the deformed magnetic fields. 
Therefore, the BZ process is explained as the zero mode limit of \alfv superradiance. 
 In this sense, \alfv superradiance is a more general energy extraction process 
 that includes the BZ process. 
Furthermore, the resonant scattering implies that \alfv superradiance can be dominant in the 
energy extraction process, although our perturbative approach will break down. 
Therefore, it is necessary to confirm this with numerical simulation.

Before closing this section, let us remark on the Kerr black hole case, 
in which there are some different points from the present model.
First, there exists outer light surface besides the inner one that is 
also causal boundary for \alf waves. Hence, we need to consider purely outgoing 
boundary conditions for \alf waves there. 
By considering the case that the \alf waves occur at an inner point of the outer light surface, where the effective potential is flat enough in the tortoise coordinate,  it is possible to use the 
same technique discussed in the present paper.
Second, the stream function $\phi_1(r,\theta)$ depends on the radial and polar coordinates. 
It makes the problem more difficult because in order to consider the force balance between 
magnetic surfaces, we need to solve the general relativistic 
Grad-Shafranov equation \cite{Blandford1977}. 
Although there are above differences, we have already confirmed that the condition 
for \alfv superradiance coincides with that for the BZ process even for the Kerr case. 
We will discuss the details in the next paper.

Moreover, when magnetic field lines connect to an accretion disk and or jet around the black hole, 
we may see interesting phenomenon: \alf waves can be confined in the finite region between the black hole and the disk or jet, then \alfv superradiance may occur repeatedly like black hole bomb \cite{Press1972a}.
%
\section{Concluding remarks}

We investigated energy extraction mechanisms from a rotating black 
cylinder spacetime with a forcefree magnetosphere to reveal the relationship 
between the BZ process and \alfv superradiance. 
Through the evaluation of the superradiant condition and the Poynting flux, 
we showed that the BZ process is, in fact, the zero mode limit of \alfv superradiance.
 The result of the present paper implies that the wave phenomenon is important for 
 discussing the engine of high-energy astrophysical compact objects such as gamma ray bursts and active galactic nuclei.
 
\begin{acknowledgements}
 The authors thank Shinji Koide for fruitful discussions. 
    Y.N. was supported in part by JSPS KAKENHI Grant No. 15K05073. 
  M.T. was supported in part by JSPS KAKENHI Grant No. 17K05439. 
  \end{acknowledgements}


%
{}

\end{document}